\documentclass[aps,10pt,prd,twocolumn,preprintnumbers,amsmath,amssymb,nofootinbib,superscriptaddress,a4paper,showpacs]{revtex4-1}

\usepackage{amsfonts}
\usepackage{mathrsfs}
\usepackage{amssymb}
\usepackage{amstext}
\usepackage{enumerate}
\usepackage{color}
\usepackage[english]{babel}
\usepackage{graphicx}% Include figure files
\usepackage{dcolumn}% Align table columns on decimal point
\usepackage{bm}% bold math
\usepackage{exscale}
\usepackage{relsize}
\usepackage{float}
\newcommand{\ignore}[1]{}
\newcommand{\Ref}[1]{(\ref{#1})}

\newcommand{\R}{{\mathbb R}}
\newcommand{\Z}{{\mathbb Z}}

\newcommand{\SU}{\mathrm{SU}}

\newcommand{\SLtc}{\mathrm{SL}(2,\mathbb{C})}

\newcommand{\h}{{\cal H}}

\newcommand{\be}{\begin{equation}}
\newcommand{\ee}{\end{equation}}
\newcommand{\bea}{\begin{eqnarray}}
\newcommand{\eea}{\end{eqnarray}}

\newcommand{\bit}{\begin{itemize}}
\newcommand{\eit}{\end{itemize}}

\newcommand{\tr}{{\rm Tr}}

\newcommand{\re}{\mathrm{Re}}

\newcommand{\ii}{\mathrm{i}}
\newcommand{\ex}{\mathrm{e}}
\newcommand{\dd}{\mathrm{d}}

\newcommand{\bra}[1]{\left\langle {#1}\right|}
\newcommand{\ket}[1]{\left|{#1}\right\rangle}

  \newcommand{\g}{\gamma}

\begin{document}

\title{Area Law from Loop Quantum Gravity}
        \author{Alioscia Hamma}
        \affiliation{Department of Physics, University of Massachusetts Boston, Massachusetts 02125-3393, USA}
        \author{Ling-Yan Hung}
        %\email{lyhung@fudan.edu.cn}
        \affiliation{Department of Physics, Fudan University, 220 Handan Road, 200433 Shanghai, China}
        \author{Antonino Marcian\`o}
        %\email{marciano@fudan.edu.cn}
        \affiliation{Department of Physics, Fudan University, 220 Handan Road, 200433 Shanghai, China}
        \author{Mingyi Zhang}
        %\email{mingyi{\_}zhang@fudan.edu.cn}
        \affiliation{Department of Physics, Fudan University, 220 Handan Road, 200433 Shanghai, China}
        \affiliation{Max-Planck-Institut f\"ur Gravitationsphysik (Albert-Einstein-Institut), Am M\"uhlenberg 1, 14476 Golm, Germany, EU}

  \date{\today}

\begin{abstract}
\noindent
We explore the constraints following from requiring the Area Law in the entanglement entropy in the context of loop quantum gravity. We find a unique solution to the single-link wave-function in the large $j$ limit, believed to be appropriate in the semi-classical limit. 
We then generalize our considerations to multi-link coherent states, and find that the area law is preserved very generically using our single-link wave-function as a building block. Finally, we develop the framework that generates families of multi-link states that preserve the area law while avoiding macroscopic entanglement, the space-time analogue of ``Schr\"{o}dinger cat''.  We note that these states, defined on a given set of graphs, are the ground states of some local Hamiltonian that can be constructed explicitly. This can potentially shed light on the construction of the appropriate Hamiltonian constraints in the LQG framework. 
    
\end{abstract}
\pacs{04.60.Pp}
\maketitle

\section{Introduction}

\noindent 
In recent years, the study of entanglement has shed new lights in the fields of quantum field theory, quantum statistical mechanics and condensed matter theory. Whenever quantum fluctuations are important, entanglement plays an important role \cite{amico}.  In the context of condensed matter physics, it has played a key role in classifying phases of matter, notably those with mass gaps, leading to the notion of topological orders \cite{Wen:1990zza,Wen:2003yv,Xiao:803748,Wen2002175} characterized by long-range entanglement \cite{Hamma200522,PhysRevA.71.022315, PhysRevLett.96.110404, PhysRevLett.96.110405}. A crucial quantifier of quantum entanglement is the entanglement entropy. For a wave-function that is the ground state of a local Hamiltonian, it is believed that the entanglement entropy should obey the so called area law \cite{eisert-arealaw}: in the limit of a large subregion $A$, its entanglement entropy is given by $S_{A}= -\tr \rho_A \ln \rho_A$, where $\rho_A$ is the reduced density matrix of subsystem $A$, and $S_{A}$ scales to leading order as the area of the boundary of $A$. 

From lessons that have been drawn from black hole entropy, which is proposed to be understood as an entanglement entropy between degrees of freedom within and outside the horizon \cite{Bombelli:1986rw,Srednicki:1993im,Frolov:1993ym}, and more recently evidence from the computation of entanglement entropy via the Ryu-Takayanagi formula  \cite{Ryu:2006bv} in the  AdS/CFT correspondence, it appears that one very important feature of quantum gravity at least in the semi-classical limit, in which some classical background geometry can be defined, is that the leading contribution to the entanglement entropy associated to a sub-region $A$ should also satisfy the area law \cite{Bianchi:2012ev}.

Loop Quantum Gravity (LQG) \cite{Rovelli:QG,ThiemannBook,Perez:LivRR} provides a perfect  arena to test the requirement of  area law, in which explicit calculations can be done. A long standing and central question in the LQG is to understand the semi-classical limit. Thus far, semi-classical states have been proposed \cite{Thiemann:2000bw, Thiemann:2000ca, Livine:CI2007, Bianchi:2009ky} in LQG, and studied in many contexts \cite{Conrady:SL2008,Barrett:2009gg,Barrett:2009mw,Han:2011re,Han:2011rf,Han:2013gna,Han:2013hna,Han:2013ina,Han:2013tap,Rovelli:PG2006,Bianchi:PGN2009,Rovelli:3P2011,Bianchi:2011hp,Riello:2013gja,Bojowald:2009jk,Bojowald:2007bg,Bianchi:2010zs, Magliaro:2010qz}. However, it is still by far an open question whether a semi-classical geometry has been undisputedly recovered. The current paper is partly inspired by recent work in LQG that recovers the Bekenstein's black hole entropy  \cite{Rovelli:1996dv, Ashtekar:1997yu, Kaul:1998xv, Ashtekar:2000eq, Domagala:2004jt, Corichi:2009wn, Engle:2009vc, bianchi2011black, Bianchi:2012ui,Han:2014xna,Achour:2014eqa} and several attempts in computing entanglement entropy in terms of spin network states \cite{livine2008bulk,livine2009entropic,Dasgupta,Donnelly:2008vx,Donnelly:2011hn,Bodendorfer:2014fua}, which is intimately related to the area law of the entanglement entropy \cite{Bianchi:2012ev,Bodendorfer:2014fua}. 

In this letter, we inspect the semi-classical limit through the lens of entanglement entropy of a bounded sub-region in LQG. We find that imposing the area law across a single triangular surface, the simplest possible scenario, already leads to a very tight constraint on the possible wave-function, which admits an almost unique solution, one that acquires an interesting correction from what is envisaged in \cite{Bianchi:2012ev}.   

Ultimately the main goal of this paper is to show that entanglement imposes very stringent requirement to a theory of quantum gravity. We show how to use this guide in a constructive way. In particular, we impose that, together with the area-law, entanglement has to obey (i) SU(2) gauge invariance, and (ii) it must be {\em microscopic}, implying that fluctuations in quantum space-times should be confined to microscopic scales. These are very natural requirements if Entanglement is physically measurable. Gauge invariance means that entanglement does result in stronger than classical correlations in actual physical observables \cite{zanardi}, and microscopic entanglement means that we do not allow arbitrarily large Schr\"{o}dinger cat-spacetimes \cite{hgi}. These considerations, together with the single-link result, allows us to construct the many-body or multi-facet state of a  general quantum geometry. As a bonus, we obtain a constructive method to build local Hamiltonians for given set of graphs, potentially important towards understanding the Hamiltonian constraints of LQG. 

\section{One-link states}
\noindent
Let us begin by setting the stage and notations. Our starting point is the $\SU(2)$ spin-network states in the graph-fixed kinematical Hilbert space $\h_\Gamma$ of LQG. An $\SU(2)$ spin-network state is a triplet 
$$\ket{\mathcal{S}}=\ket{\Gamma, j_l, i_n}\,.$$ 
$\Gamma$ is the given \emph{proper} graph with $L$ oriented links and $N$ nodes. $j_l\in\Z^+/2$ is an assignment of an $\SU(2)$ unitary irreducible representation to each link $l$, and $i_n$ is an assignment of an $\SU(2)$ intertwiner to each node $n$. The spin-$j_l$ representation is isomorphic to a $\SLtc$ representation $(\gamma j_l,j_l)$ selected by the linear version of the simplicity constraints \cite{Engle:2007uq}, namely 
$$\vec{K}=\gamma\vec{L}\,,$$ 
in which $\vec{K}$ and $\vec{L}$ are boost and rotation generators, respectively. The simplicity constraints are used in the spin foam formalism \cite{Engle:2007wy,Freidel:FK2008,Kaminski:2009fm,Speziale:2012nu} for recovering gravity amplitudes from BF(topological) amplitudes\cite{Plebanski:1977zz,Capovilla:1991qb} and for regaining the real connection \cite{Wieland:2010ec,Speziale:2012nu}. Spin-network states correspond to the so called \emph{twisted geometries} \cite{Freidel:2010aq,Freidel:2010bw}, which describe the geometries of three dimensional \emph{fuzzy} discrete manifolds. Each $N$-valent node corresponds to a $N$-facet polyhedron, while each link is dual to the face of the polyhedron \cite{Bianchi:2010gc}. The areas of the faces are realized as the expectation value of the SU(2) Casimir operator acting on the spin-network states, {\it i.e.} the area of the face dual to link $l$ is 
\begin{equation}\label{eq:Al}
\langle\hat{\mathcal{A}}_l\rangle=8\pi\gamma \ell_p^2\sqrt{j_l(j_l+1)},
\end{equation}
where $\gamma\in\R^+$ in the rest of the letter, and $\ell_p$ is the Planck length.
\begin{figure}[htbp!]
\centering
\includegraphics[width=.45\textwidth]{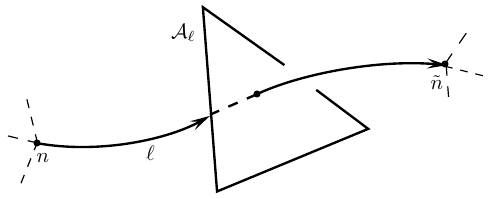}
\caption{\label{fig:Al}A face $\mathcal{A}_l$ dual to link $l$.}
\end{figure}

\noindent 
Given a state $\ket{\Psi}$ in $\h_\Gamma$ and cut the graph $\Gamma$ into two regions $A$ and $\bar{A}$, the entanglement entropy $S_A$ between these two regions is defined as the von Neumann entropy of the reduced density matrix  
$$\rho_A\equiv \tr_{\bar{A}}\ket{\Psi}\bra{\Psi}\,.$$ 
$\tr_{\bar{A}}$ is the partial trace over the states ({\it i.e.} the intertwiners' and spins' degrees of freedom) inside $A$.  To recover locally a Minkowskian vacuum that respects the Bisognano-Wichmann theorem,  it is proposed that the reduced density matrix $\rho_A$ should be proportional to the exponentiation of the boost operator $K$, leading to the following proposal for a single-link state \cite{Bianchi:2012ev,Chirco:2014naa}
$$\vert {\Psi} \rangle= \sum_j\alpha_j\exp(- \gamma K)_{mn}\vert j,m,n\rangle\,.$$ 
A careful check however suggests that it \emph{does not} satisfy the area law %\cite{longpaper} 
in the large $j$ limit, which is another crucial ingredient of the semi-classical limit, opposing naive expectation. Indeed, one can readily check that the normalization of the reduced density matrix always cancel the area term in this limit.  Surprisingly, in the large $j$ regime, there is a natural and arguably unique solution that recovers the area law across each link.

By construction, we will implement invariance under gauge and space-diffeomorphisms of the states $|\Psi\rangle$ that we seek to fulfill the area law requirement. More in general, any reduced density matrix that is related by unitary transformation to the density matrix associated to the $|\Psi\rangle$ state will have the entanglement entropy scaling as area, because of the expressions of the entanglement entropy involved. This ensures that our procedure is effective in order to determine a generic state $|\Psi\rangle$ with area law scaling in the large $j$ limit, which is unique up to unitary transformations. At this purpose, we first recover in this section the single-link state that achieves this goal, then generalize in the next sections our proposal to multi-link coherent states and to multi-nodes states, so to fully encode gauge and space-diffeomorphism invariance in the construction.

For a single-link we can conveniently choose a state in the diagonalized basis with fixed spin $j$. This is expressed by  
\begin{eqnarray}\nonumber
|\Psi \rangle= \frac{1}{\mathcal{N}} \, \sum_{m, n=-j}^j F_{m n}^{j}\, |j,m,n \rangle =   \frac{1}{\mathcal{N}} \, \sum_{m=-j}^j \ F_{m m}^{j}\, |j,m \rangle \,,
\end{eqnarray}
having used the form of the coefficients in the diagonalized basis, i.e. $F_{m n}^j=F_{m m}^j\, \delta_{m n}$ and $ |j,m \rangle\equiv |j,m,m \rangle$.
The reduced density matrix casts 
\begin{eqnarray} \nonumber
\rho_A \equiv \frac{1}{\mathcal{N}} \, \sum_{n=-j}^j \overline{F}_{n m} \, F_{n m'} \, \ket{j,m,n}\!\bra{j,m',n}, 
\quad \,\,
\end{eqnarray}
and because of the diagonalized basis involved, satisfies
\begin{eqnarray} \label{ra}
&&\rho_A=\! \sum_{m=-j}^j p(m) \ket{j,m}\!\bra{j,m},  \\
&&p(m)\equiv\frac{f(m)}{\sum_{m=-j}^j f(m)}, \nonumber
\end{eqnarray}
with $f(m)= \overline{F}_{m m} \, F_{m m} $. Note that the function $f(m)$ is assumed not to carry explicit $j$ dependence. \\

In what follows, by determining the form of $f(m)$, we show that 
\begin{eqnarray}\nonumber
F^j_{m n}= D_{mn}^{j}\left(%g\, 
\ex^{-\pi\gamma L_{z}
%+\ii\phi_l L_z
-\frac{\exp(1-2\pi \gamma L_z)}{4\pi\gamma}}%\tilde{g}^{\dag}
\right),
\end{eqnarray}
where $D^{j}_{mn}(\cdots)$ is defined as $\bra{j,m}\cdots\ket{j,n}$ %, $g_l$ and $\tilde{g}_l$ denotes SU(2) elements, 
and $L_z$ represents the $\mathfrak{su}$(2) generator in the $z$ axis. In the next sections, we will use this single-link state as the building block to construct generalized multi-link and multi-node states. 

We now focus on \eqref{ra}. 
Since we are interested in the large spin regime, the sum can be replaced by an integral 
$$\sum_{m=-j}^j \rightarrow \int_{-j}^jdj\,. $$ 
Requiring that the entropy admits an area law and taking into account \Ref{eq:Al}, we impose 
$$S_A = c j + \cdots\,,$$ 
the derivative of which w.r.t. $j$ is thus a constant $c>0$. This leads to a constraint on $f(m)$ \footnote{We have made the simplifying assumption that the probability density for $m<0$ is negligible compared to $m>0$. Relaxing this assumption does not lead to any material change. }:
\be \label{pj}
\frac{c}{p(j)}+ c j -1 + \ln p(j) =0.
\ee 
Eq. (\ref{pj}) cannot be solved exactly in complete generality. However, anticipating that we are considering the large spin limit, and motivated by the original proposal, we consider the limit in which  
$$\frac{c}{p(j)}\ll |cj-1|\,.$$
In this limit, we have $p(j) = \exp(1-cj)$, which finally gives
\be \label{eq:prob}
f(m)= \exp\left[{- c\, m  - \frac{\exp(1 -  c\, m)}{c}}\right].
\ee  
One can readily check that ${c}/{p(j)}\ll |cj-1|$ is satisfied in the large $j$ limit, namely { given $J\in \mathbb{R}^+$ in the limit $J \ll j$, if $c$ satisfies}
\begin{equation}\label{eq:limit}
0<J^{-1}\ll c \ll J^{-\frac{1}{2}}\ll 1\,.
\end{equation}
In this limit the leading term in the entanglement entropy is indeed linear in $c j$, recovering the long sought area law. The expression for $c$ is then straightforwardly determined to be $c=2\pi \gamma$. 

We remark that the area law is indeed an upper bound for the entanglement reachable by these states. In particular, there are non-logarithmic corrections scaling like the volume of the system. One consequence is that these states cannot be seen even locally as finite temperature equilibrium states. In quantum cosmology, if one sought to find reduced density matrices that represent thermal states, these should be obtained by tracing out high energy states. The ground state --- or the low lying states --- will only feature the area law. On the other hand, these states can be topologically ordered. Topological order does indeed consist in a global constraint on the allowed configurations on the boundary, resulting in a negative correction to the entanglement entropy \cite{PhysRevA.71.022315, PhysRevLett.96.110404, PhysRevLett.96.110405,renyi}.

The  function $p(m)$ also naturally suppresses contributions at small values of $m$ in this limit, justifying the approximation that replaces the sum by the integral. It is very interesting that the state that we have found is very similar to the original proposal, up to an exponentially suppressed factor, which is negligible where $f(m)$ actually contributes, but suppresses regions which would otherwise have contributed in the original proposal. The constant $c$ also emerges naturally and plays precisely the same role as the Barbero-Immirzi parameter $\gamma$, and \Ref{eq:limit} is exactly the semi-classical large $j$ regime in the covariant formalism of LQG \cite{Han:2013tap}. 
Recall that the boost operator $K_z$ was related to $L_z$ via the simplicity constraint 
$$\vec{K}=\gamma\vec{L}\,,$$ 
which followed from an action quadratic in the tetrad. One might be tempted to interpret the extra exponential correction we find here as a non-linear correction  arising from quantum effects. 

We can extend our construction to the more complicated case of one node, with many links protruding from it. For simplicity, the graph $\Gamma$ used in the following calculation is the graph with $L$ out-pointing links attaching to only one node.
\begin{figure}[htbp!]
\centering
\includegraphics[width=.4\textwidth]{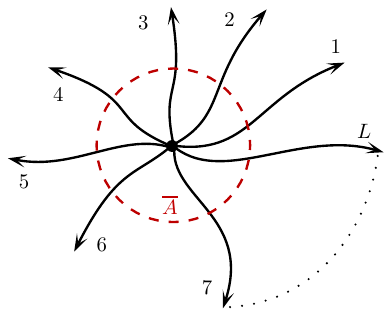}
\caption{\label{fig:graph}A graph with $L$ links and one node. }
\end{figure}

\section{Multi-link states}
\noindent
Combining our {previous one-link} proposal with the well known construction of multiple link coherent spin-network state \cite{Thiemann:2000bw, Thiemann:2000ca, Livine:CI2007, Bianchi:2009ky}, we are led to consider the state $\ket{\Psi}$ supported on $\h_\Gamma$, which reads
\begin{eqnarray} \label{onenode}
|\Psi \rangle &=&\frac{1}{\mathcal{N}}\sum_{\left\{ j_{l}\right\} }\int_{SU\left(
2\right) }dh\left( \prod_{l}^{L}\Delta^{j_l}_{(t_l,J_l)}\right)\times\\
&&\times \sum_{\left\{ m_{l}\right\} \left\{ k_{l}\right\}
}\sum_{\left\{ n_{l}\right\}
}\prod_{l}^{L}F_{m_{l}n_{l}}^{j_{l}}D_{n_{l}k_{l}}^{j_{l}}\left( h\right)
|j_{l},m_{l},k_{l}^{\dag }\rangle. \nonumber
\end{eqnarray}
In \eqref{onenode} we defined: the normalization factor $\mathcal{N}$, which is recovered in Sec.~\ref{ap.A} of the appendix; 
$$\Delta^{j_l}_{(t_l,J_l)}\equiv d_{j_l}\ex^{-(j_l-J_l)^2t_l}\,$$ 
with $d_{j_l}=2j_l+1$; $J_l$, a given spin for the link $l$, and $t_l$, similar to a heat kernel time; the Wigner matrix of the group elements that encodes the correction found in \eqref{eq:prob}, i.e.
\begin{equation}\label{eq:F}
F^{j_l}\equiv D^{j_l}\left(g_{l}\ex^{-\pi\gamma L_{z}+\ii\phi_l L_z-\frac{\exp(1-2\pi \gamma L_z)}{4\pi\gamma}}\tilde{g}_{l}^{\dag}\right),
\end{equation}
in which $D^{j}_{mn}(\cdots)$ is defined as $\bra{j,m}\cdots\ket{j,n}$; $\phi_l$ is a phase; $g_l$ and $\tilde{g}_l$ are SU(2) elements; $L_z$ is the $\mathfrak{su}$(2) generator in the $z$ axis; $v_{i}$ is the normalized intertwiner with 
$$\overline{v}_{i}\cdot v_{i'}=\delta_{ii'}\,.$$ 
Note that this is a state that is factorizable into wave-functions of individual links up to gauge constraints. \\

We cut the graph into two regions as in Fig.~\ref{fig:graph}. Region $\bar{A}$ contains the node. The reduced density matrix $\rho_A$ is recovered by tracing out the link source states (or the intertwiner degrees of freedom) in $\ket{\Psi}$. Furthermore, $\rho_A$ can be rewritten in the block-diagonal form 
$$\rho_A=\sum_{\{j_l\}}\rho_{\{j_l\}} = \sum_{\{j_l\}}P_{\{j_l\}}\hat{\rho}_{\{j_l\}}\,,$$ 
where each block is characterized by the configuration of the $\{j_l\}$ eigenvalues of the boundary links. Crucially, such a block factorization follows from the SU(2) invariance imposed at the node and the properties of the intertwiners. We can write $P_{\{j_l\}}$, the norm of the $\{j_l\}$'s block $\rho_{\{j_l\}}$, as
\begin{equation}\label{eq:Pj}
P_{\{j_l\}}\equiv\frac{1}{\mathcal{N}^2}\int_{\SU(2)} \dd h\prod_l^L[\Delta^{j_l}_{(t_l,J_l)}]^2\chi_{j_l}[\tilde{g}_{l}f(L_z)\tilde{g}_{l}^{\dag}h].
\end{equation}
In \eqref{eq:Pj} we use that: $f(L_z)$ is given by \Ref{eq:prob} with $c\equiv 2\pi\gamma$; $\chi_{j_l}(M)$ is the trace over matrix $M$; $\hat{\rho}_{\{j_l\}}$ is the reduced density matrix with fixed $\{j_l\}$
\begin{equation}
\hat{\rho}_{\{j_l\}}\equiv\frac{1}{Z_{\{j_l\}}}\int_{\SU(2)}\dd h\prod_l^L F^{j_l}\cdot D(h)\cdot F^{j_l\dag},
\end{equation}
where $F^{j_l}$ is given by \Ref{eq:F}; $Z_{\{j_l\}}$ is the normalization such that $\tr\hat{\rho}=1$. The integration over $\SU(2)$ element $h$ follows from the summation of intertwiner, which is proportional to an integration over Wigner matrices on SU(2) 
$$\sum_i\overline{v_i^{\{p_l\}}}v_i^{\{p'_l\}}\propto \int_{\SU(2)}\dd h \prod_l^LD_{p_lp'_l}^{j_l}(h)\,.$$ 
The normalization has been absorbed into $\mathcal{N}$. Then the entanglement entropy 
$$S_A=-\tr\rho_A\ln\rho_A$$ 
evaluates to
\begin{eqnarray}\label{eq:vonNeumann}
S_A= -\sum_{\{j_l\}}P_{\{j_l\}}\tr\hat{\rho}_{\{j_l\}}\ln\hat{\rho}_{\{j_l\}}-\sum_{\{j_l\}}P_{\{j_l\}}\ln P_{\{j_l\}}\nonumber \,.
\end{eqnarray}
Note that:\\

i) $P_{\{j_l\}}$ can be understood as the probability density of detecting the given boundary configuration $\{j_l\}$, and that 
$$\sum_{\{j_l\}}P_{\{j_l\}}=1\,;$$ 

ii) it is a subtle issue to define the entanglement entropy in a gauge theory \cite{Buividovich:2008gq,Casini:2013rba} --- moving from the proposal in \cite{Bianchi:2012ev}, the resultant $S_A$ is in fact a SU(2) gauge invariant quantity which only has dependence on the eigenvalues of the Casimir of the boundary links, a choice equivalent to the proposal in \cite{Hung:2015fla}; \\

iii) the entanglement entropy with a unique boundary link-configuration is given in the first term of \Ref{eq:vonNeumann}, and it will be proved later to be the area term in the large spin limit. \\

To calculate $S_A$, we first give an explicit expression for $P_{\{j_l\}}$ and $\tr\hat{\rho}_{\{j_l\}}\ln\hat{\rho}_{\{j_l\}}$. As such, we introduce the coherent states on SU(2) \cite{Livine:CI2007}, namely 
$$D^j_{mj}(n)\ket{j,m}=n\ket{j,j}=n\ket{\uparrow}^{\otimes 2j}\,,$$ 
with $n\in\SU(2)$. Imposing the $\SU(2)$ coherent states resolution of identity into each $\chi_{j_l}$ in \Ref{eq:Pj}, $P_{\{j_l\}}$ becomes,
\begin{eqnarray} \nonumber 
P_{\{j_l\}}=\frac{1}{\mathcal{N}^2}\int\dd h\prod_l^L\dd n_l~[\Delta^{t_l}_{(j_l,J_l)}]^2 d_{j_l}\sum_{p_l}^{\infty}\frac{(-\ex)^{p_l}}{p_l!(2\pi\g)^{p_l}}\ex^{S_{p}},
\end{eqnarray}
where the ``action'' $S_p$ is given by
\begin{eqnarray}
S_{p}\equiv \sum_l^L2j_l\ln\bra{\uparrow}n^{\dag}_l\tilde{g}_{l}e^{-2\pi\g (1+p_l)L_{z}}\tilde{g}_{l}^{\dag}h~ n_l\ket{\uparrow}\,.
\end{eqnarray}
In order to study the large spin behavior of $P_{\{j_l\}}$, we rescale $j_l\rightarrow \lambda k_l$ and assume $\lambda\gg 1$. It is convenient to perform the asymptotic expansion to get the major contribution of $P_{\{j_l\}}$. The solutions of the equations of motion 
$\delta _{n_{l}}S_p=\delta _{h}S_p=0$ 
control the semi-classical behavior of $P_{\{j_l\}}$. For $\delta _{n_{l}}S_p=0$, the only possible solutions \footnote{Another solution $-\epsilon \bar{n}_{l} = \tilde{g}_{l}, \re(\alpha _{l}) =-2\pi\gamma$ will lead $P_{\{j_l\}}\propto \exp(-\ex^{2\pi\g j})$ and suppressed faster than the solution $n_{l} = \tilde{g}_{l}, \re(\alpha _{l}) =2\pi\gamma$ in the large spin limit.} are
$n_{l} = \tilde{g}_{l},\quad h =\mathbf{1}$. 
Note that $\delta _{h}S_p=0$ is equivalent to the Gauss constraint. Then $P_{\{j_l\}}$, to leading order of the asymptotic expansion and in the limit of \Ref{eq:limit}, becomes
\begin{equation}\label{eq:Pja}
P_{\{j_l\}}=\frac{1}{\mathcal{N}^2}\prod_l^L~[\Delta^{t_l}_{(j_l,J_l)}]^2\frac{1}{j^{3/2}}\exp\left(-1-\frac{\ex^{1-2\pi\g j}}{2\pi\g}\right).
\end{equation}
where $j$ is the average of $j_l$. 
In order to estimate the term $\tr\hat{\rho}_{\{j_l\}}\ln\hat{\rho}_{\{j_l\}}$, it is easier to use the replica trick to compute $S_n$, the R{\'e}nyi entropy of order $n$~\cite{1742-5468-2004-06-P06002}. From 
$$\tr\hat{\rho}_{\{j_l\}}^n=\prod_l^L \ex^{(n-1)(1-2\pi\g j_l)}$$ 
and  taking the $n\rightarrow 1$ limit, we get
$$
-\tr\hat{\rho}_{\{j_l\}}\ln\hat{\rho}_{\{j_l\}}=\sum_l^L 2\pi\g j_l-3/2\ln\sum_l j_l^2\,.
$$
The logarithmic terms is from the gauge invariance \cite{bianchi2011black}.  Finally, we have

\noindent
\begin{equation}\label{eq:Sa}
%\begin{split}
S_A=%&
\sum_l^L\langle 2\pi\g j_l\rangle-\frac{3}{2}\left\langle\ln\sum_l j_l^2\right\rangle-\langle\ln P_{\{j_l\}}\rangle\,,
%\end{split}
\end{equation}
where we have denoted 
$$\langle \cdots\rangle\equiv \sum_{\{j\}}(\cdots)P_{\{j\}}\,.$$
The first term is the area $\mathcal{A}$ of region $A$'s boundary. The second term has contribution which are proportional to link number $L$. Disregarding terms independent from the boundary area, when $t=J^{-k}$ the logarithmic correction amounts to $-3k/2\ln(\mathcal{A}/\ell^2_p) $. So in the semi-classical limit of LQG \Ref{eq:limit}, the entanglement entropy is
\begin{equation}
S_A=\frac{\mathcal{A}}{4\ell_p^2}+\mu L-\frac{3k}{2}\ln\frac{\mathcal{A}}{\ell^2_p}\,,
\end{equation}
where the ``chemical potential'' term now carries only $1/j$ suppressed contributions.

\section{Multi-node states}
\noindent
At this point, we generalize to multi-node states. In the following, we show how we can produce a class of many-body states that satisfy all the requirements of area law and gauge invariance spelled above, and that can be the ground states of a local Hamiltonian. 

Let us start with the simples possible multi-node state obtained by considering the tensor product of single-node states over all the nodes of the graph and then summing over all the $\{j_l\}, \{i_n\}$ in order to enforce gauge invariance. This state reads 
\begin{eqnarray} \label{mb}
\ket{\Psi}&\equiv&\frac{1}{\mathcal{N}}\sum_{{\{j_l\},\{i_n\}}}\prod_{l,n}\Delta^{j_l}_{(t_l,J_l)}F^{j_l}\cdot\overline{v}_{i_n}\ket{\Gamma, j_l, i_n} \nonumber\\
&=&\frac{1}{\mathcal{N}}\sum_{{\{j_l\},\{i_n\}}}\prod_{l,n} |\varphi\rangle_{n,l}\,.
\end{eqnarray}
The above many-body state still satisfies the area law, because, although the number of nodes and links contained in  $\bar A$  scales with its volume, however links that are completely enclosed within the region $\bar A$ would not contribute to the entanglement entropy, which will in stead scale with the number of links crossing the boundary $\partial \bar A$ yielding  the area law,  
$$\mathcal A = \sum_{l\in \partial \bar A}\langle \hat{\mathcal A_l}\rangle\,.$$ 
This statement is confirmed by a series of results within LQG that describe the space of black hole microstates --- see e.g. Ref.~\cite{Engle:2009vc} and formerly Ref.~\cite{Smolin}. In particular, in Ref. \cite{Krasnov:2009pd} was argued that for an observer at infinity a black hole can be described by an SU(2) intertwining operator, effectively corresponding to a ``gigantic'' single tensor operator intertwining all the links that puncture its horizon. This procedure corresponds to the large $j$ limit we are interested in for this analysis\footnote{We thank the Referee for whipping this clarification up.}.\\
From the many-body point of view, this state is trivial, as there are no quantum fluctuations other than the single-body ones. However, exactly because it is trivial, it is easy to build a local quantum Hamiltonian such that the state in Eq.(\ref{mb}) is its ground state. Let $\pi$ be the projector onto the gauge-invariant Hilbert space and $\pi_{n,l} =|\varphi\rangle\langle\varphi\rangle |_{n,l}$ the projector onto the local state $|\varphi\rangle_{n,l}$. Then the Hamiltonian 
$$H(0)=-\pi(\sum_{n,l} \pi_{nl})\pi^\dagger$$ 
is a local Hamiltonian whose ground state is state $|\Psi\rangle$ in Eq.(\ref{mb}).  \\
 
Moreover, one has to admit that the ansatz Eq.(\ref{mb}), although motivated by the requirements of entanglement and gauge invariance discussed so far, still seems very arbitrary and fine tuned. We have no reason to care about a particular kinematical state, after all. What counts, are the physical states. And yet, physical states should be in the same 'equivalence class' of the kinematical state defined above, in the sense that they must retain all the properties we want. Unfortunately, one cannot simply preserve these properties by just writing any state in the kinematical subspace. Indeed, 
 a generic superposition of such kind of states, though, will first of all not respect the area law, and will contain arbitrarily macroscopic superpositions. 
 
In order to form the right equivalence class, we need to deform the trivial state in a way that some many-body entanglement and quantum fluctuations are produced, but without violating the constraints we set up. This can be obtained by the technique of {\em quasi-adiabatic continuation} introduced in \cite{quasiadiabatic}. In this way, a whole class of states can be defined, that constitute what is known in condensed matter theory as a quantum phase. This technique allows to continuously deform a quantum state that is the ground state of a local Hamiltonian in order that it is still the ground state of a local Hamiltonian. Let $ H(\lambda)$ be a smooth family of {\em local} Hamiltonians parametrized by  $\lambda\in [0,1]$ such that 
$$\ket{\Psi_N} = \ket{\Psi(0)}$$ 
is the ground state of $H(\lambda=0)$ just like in  Eq.(\ref{mb}). Here, local means that  $H(\lambda)$ is the sum of local operators, 
$$H(\lambda) = \sum_X \hat T_{X\subset\Gamma} (\lambda),$$ 
where $X$ denotes a subset of the graph $\Gamma$ and $\hat T_X (\lambda)$ means that this operator has only support on $X$. Also assume that $H(\lambda)$ has a finite gap $\Delta E$ between ground and first excited state for all $\lambda$'s. One way to think of $H(\lambda)$ is as the perturbation of the initial Hamiltonian $H(0)$, namely, 
$$H(\lambda) = H(0) + \lambda \sum_X \hat{K}_X$$ 
and $\hat{K}_X$ are any local operators with support on $X$. Then, following \cite{bravyi}, one can define a unitary operator 
$$U(\lambda) = \mathcal T \exp \{ -i \int^\lambda _0 \tilde{H}(s)ds\}$$ 
with $\mathcal T$  the time ordering operator, and 
$$\tilde{H}(s) = i\int dt F(t) \, e^{i H(s) t}\, \partial_s H(s) \, e^{-iH(s)t}\,,$$
with $F(t)$ an appropriate fast decaying smooth function --- see \cite{bravyi} for details. The unitary operator $U(\lambda)$ so defined interpolates among ground states of $H(\lambda)$ and it is therefore called { adiabatic continuation}  \cite{quasiadiabatic}. The ground state of $H(\lambda)$, $\ket{\Psi(\lambda)}$, can be written as the adiabatic continuation of $\ket{\Psi(0)}$, namely 
$$\ket{\Psi(\lambda)}=U(\lambda)\ket{\Psi(0)}\,.$$ 
Now, the adiabatic continuation preserves the area law  \cite{eisert-arealaw}. In this way, we can always deform $\ket{\Psi(0)}$ in a way to obtain a new state $\ket{\Psi(\lambda)}$ such that its parent Hamiltonian $H(\lambda)$ is non-integrable. The above definition is constructive as for any choice of the $\hat{K}_X$ we can construct a family of local Hamiltonians and their gauge-invariant, area-law ground states.

As an application of this method, let us show that the so defined states do not have macroscopic superpositions, like macroscopic space-time Schr\"odinger cat states. The initial state $|\Psi\rangle$ is by construction not macroscopically entangled. We now show that by adiabatic continuation we also preserve the physicality of entanglement as being microscopic. Adiabatic continuation preservers macroscopic entanglement. So, if one continues a state that is not a Schr\"odingers' cat, one will not obtain a Schr\"odingers' cat. On the other hand, 
continuing a Schr\"odingers' cat will still yield a macroscopically entangled state. To show this, we apply a result obtained in \cite{hgi}. The superselection rule is imposed by stating that states with non vanishing mutual information $\mathcal I_\infty$  between two distant macroscopic regions must be ruled out.
The mutual information is defined as 
$$\mathcal I(A|B) := S(A) + S(B) - S(AB)\,.$$ 
Here, $S$ is chosen to be the $2-$R\'enyi entropy 
$$S_2 = -\log \mbox{Tr} \rho_A^2\,,$$
namely, the logarithm of the purity instead the von Neumann entropy. The technical reason for making this choice is that one can prove that $\mathcal I_\infty$ is preserved by quasi adiabatic continuation. The physical reason, is that this quantity is a physical observable (unlike the Von Neumann Entropy) \cite{ekert2002direct,abanin2012measuring,daley2012measuring}, but still captures all the entanglement properties that are important in quantum many-body theory \cite{renyi, Isakov}.   As a bonus, starting with a trivial non-interacting Hamiltonian that is the sum of commuting local terms in the tensor product structure of Eq.(\ref{mb}), one can adiabatically continue the very Hamiltonian by means of any other Hamiltonian with a gap, which is the sum of local operators with couplings depending on some parameter $\lambda$. This is thus a systematic construction of local Hamiltonians whose ground states satisfy the desired properties we have set forth. 

\section{Summary \& outlook}
\noindent
To conclude, we have shown that entanglement can provide stringent guiding principles in selecting states with sensible semiclassical limits in LQG. We obtain these results by imposing that entanglement must be physical (gauge invariant and not macroscopic), deploying methods from quantum information and quantum many-body theory. In perspective, the framework here developed will allow for the study of notions like quantum order or thermalization in a closed quantum system in the context of quantum gravity. 

\acknowledgments 
\noindent 
We wish to thank the referee for whipping a clarification up in Sec.~IV about the inclusion of multi-nodes states in our analysis. 
This work was supported in part by the National Basic Research Program of China Grant 2011CBA00300, 2011CBA00301 the National Natural Science Foundation of China Grant  61033001, 61361136003.

\appendix 

\section{The entanglement between half links}
\noindent 
In this section of the appendix we provide an asymptotic analysis of the one-link state entanglement calculated between its two half links. We start by first checking the asymptotic analysis of the normalization of the one-link state. \\

The normalization of the state can be written as
\begin{eqnarray*}
\mathcal{N}^2\equiv I&=&d_{j}\int dn\langle j,j|n^{\dag }e^{-cL_{z}-\frac{\exp \left(
1-cL_{z}\right) }{c}}n|j,j\rangle  \\
&=&d_{j}\int dn\langle j,j|n^{\dag }e^{-cL_{z}}\sum_{k}^{\infty }\frac{%
\left( -\right) ^{k}}{k!c^{k}}e^{k[\left( 1-cL_{z}\right) ]}n|j,j\rangle  \\
&=&d_{j}\sum_{k}^{\infty }\frac{\left( -\right) ^{k}}{k!c^{k}}\int dn\langle
j,j|n^{\dag }e^{-cL_{z}}e^{k[\left( 1-cL_{z}\right) ]}n|j,j\rangle  \\
&=&d_{j}\sum_{k}^{\infty }\frac{\left( -\right) ^{k}e^{k}}{k!c^{k}}\int
dne^{S_{k}}\,,
\end{eqnarray*}%
where%
\begin{equation*}
S_{k}=2j\ln \langle \uparrow |n^{\dag }e^{-c\left( 1+k\right)
L_{z}}n|\uparrow \rangle\, .
\end{equation*}
The latter is the ``action" defined in eq.(10) of the manuscript. We search the saddle point of $S_{k}$ by variating it with respect to $n$, and obtain
\begin{eqnarray*}
\delta _{n}S_{k}&=&2j\frac{\bar{\eta}\langle \downarrow |n^{\dag }e^{-c\left(
1+k\right) L_{z}}n|\uparrow \rangle +\eta \langle \uparrow |n^{\dag
}e^{-c\left( 1+k\right) L_{z}}n|\downarrow \rangle }{\langle \uparrow
|n^{\dag }e^{-c\left( 1+k\right) L_{z}}n|\uparrow \rangle }\nonumber\\
&=&0\,.
\end{eqnarray*}
The latter relation leads to 
\begin{equation} \label{n}
n^{\dag }e^{-c\left( 1+k\right) L_{z}}n=e^{-\alpha L_{z}}\,,
\end{equation}%
with $\alpha $ a complex number. Equation (\ref{n}) has two solutions: one is 
\begin{equation*}
nL_{z}n^{\dag }=L_{z},\quad \alpha ^{+}=c\left( 1+k\right)\,; 
\end{equation*}%
the other one is 
\begin{equation*}
nL_{z}n^{\dag }=-L_{z},\quad \alpha ^{-}=-c\left( 1+k\right) \,.
\end{equation*}%
On the saddle point, we then obtain
\begin{equation*}
S_{k}^{\pm }=\mp c\left( 1+k\right) j \,.
\end{equation*}
One of the leading contribution to the normalization $I$ is given by
\begin{eqnarray*}
&&d_{j}\sum_{\epsilon =\pm }e^{-\epsilon cj}\sum_{k}^{\infty }\frac{\left(
-\right) ^{k}}{k!c^{k}}e^{\left( 1-\epsilon cj\right) k} \\
&=&d_{j}\sum_{\epsilon =\pm }\exp \left( -\epsilon cj-\frac{e^{\left(
1-\epsilon cj\right) }}{c}\right) \equiv d_{j}\sum_{\epsilon =\pm
}e^{S_{0}^{\epsilon }}\,.
\end{eqnarray*}
Another leading contribution comes from the Hessian. Indeed, the total leading contribution of $I$ can be written as
\begin{equation*}
I\sim d_{j}\sum_{\epsilon =\pm }\frac{e^{S_{0}^{\epsilon }}}{\sqrt{\det
H^{\epsilon }}}\equiv \sum_{\epsilon =\pm }I^{\epsilon }\,.
\end{equation*}
In order to compute the Hessian, we introduce the variation of $n$ on a spin 
$j$ representation:        
\begin{eqnarray*}
\delta _{n}n|j,j\rangle  &=&\delta _{n}|j,n\rangle  \\
&=&\delta _{n}|n\rangle ^{2j} \\
&=&\eta \sum_{i=1}^{2j}n\rhd |\uparrow \cdots \downarrow _{i}\cdots \uparrow
\rangle  \\
&=&\eta \sqrt{2j}n|j,j-1\rangle  \,.
\end{eqnarray*}
The total ``action" now reads
\begin{equation*}
S=\ln \langle j,j|n^{\dag }e^{-cL_{z}-\frac{e\exp \left( -cL_{z}\right) }{c}%
}n|j,j\rangle \,.
\end{equation*}
Moving from its definition, the Hessian is found to be
\begin{eqnarray*}
&&-\frac{1}{2}\eta _{a}H_{ab}^{\epsilon }\eta _{b} \equiv -\frac{1}{2}\delta
_{n}^{2}S|_{\epsilon } \\
&&=-2j\frac{-\bar{\eta}\eta e^{-\epsilon cj-\frac{e\exp \left( -\epsilon
cj\right) }{c}}+\bar{\eta}\eta e^{-\epsilon c\left( j-1\right) -\frac{e\exp
\left( -\epsilon c\left( j-1\right) \right) }{c}}}{e^{-\epsilon cj-\frac{%
e\exp \left( -\epsilon cj\right) }{c}}} \\
&&\sim 2j\bar{\eta}\eta \left( -\epsilon c+\exp \left( 1-\epsilon cj\right)
\right) \,.
\end{eqnarray*}
Consequently, the determinant of the Hessian can be expressed in following way:
\begin{equation*}
\det H^{\epsilon }=j^{2}\left( -\epsilon c+e^{ 1-\epsilon cj} \right)^{2} \,.
\end{equation*}
For $\epsilon =1$ and within the semiclassical limit we are considered in eq.(4) of the manuscript, we find  
\begin{equation} \label{c}
c\sim\alpha e^{ 1-cj} \,. 
\end{equation}
In equation (\ref{c}), we are assuming $\alpha$ to be $O\left( 1\right)$, and at the same time $\alpha>1$. 
Then we find
\begin{equation*}
\det H^{+}\sim j^{2}\left[ \left( 1-\alpha \right) e^{ 1-cj} \right]^{2},
\end{equation*}
which entails for $I^{+}$ the following expression:
\begin{equation*}
I^{+}\sim \frac{d_{j}}{j\left( \alpha -1\right) }\exp \left( -1-\frac{e^{\left( 1-cj\right) }}{c}\right) \,.
\end{equation*}
For $\epsilon =-1$, we find
\begin{equation*}
\det H^{-}\sim j^{2}\left( \exp \left( 1+cj\right) \right) ^{2}\,,
\end{equation*}
and finally
\begin{equation*}
I^{-}\sim \frac{d_{j}}{j}\exp \left( -1-\frac{e^{\left( 1+cj\right) }}{c}
\right) \,.
\end{equation*}
It is obvious that in the semi-classical limit it holds 
\begin{equation*}
I^{+}>\!\!>I^{-}\,.
\end{equation*}
Thus the leading contribution of the normalization reads
\begin{equation*}
I\sim I^{+}\sim \exp \left( -1-\frac{e^{\left( 1-cj\right) }}{c}\right)\,, 
\end{equation*}
which coincides with the explicit calculation. For one-link the reduced
density matrix $\rho _{j}$ is found to be 
\begin{equation*}
\rho _{j}=\frac{1}{I}\sum_{m}e^{-cm-\frac{\exp \left( 1-cm\right) }{c}
}|j,m\rangle \langle j,m| \,.
\end{equation*}
In order to calculate the entanglement entropy, we use the replica trick. We then consider 
\begin{equation*}
S_{EE}=\lim_{N\rightarrow 1}\frac{1}{1-N}\ln tr\rho _{j}^{N}\,.
\end{equation*}
The trace of $\rho _{j}^{N}$ is found to be
\begin{equation*}
tr\rho _{j}^{N}=\frac{1}{I^{N}}\chi _{j}\left( e^{-Ncm-N\frac{\exp \left(
1-cm\right) }{c}}\right) \,.
\end{equation*}
Using the same asymptotic analysis we have performed above, it is straightforward to see that 
\begin{equation*}
\chi _{j}\left( e^{-Ncm-N\frac{\exp \left( 1-cm\right) }{c}}\right) \sim
\frac{1}{N}e^{-1+\left( 1-N\right) cj-N\frac{\exp \left( 1-cj\right) }{c}}\,.
\end{equation*}
Therefore, we can now calculate the trace of $\rho _{j}^{N}$ in the asymptotic limit, and find 
\begin{eqnarray*}
tr\rho _{j}^{N} &\sim &\frac{\exp \left( -1+\left( 1-N\right) cj-N\frac{\exp
\left( 1-cj\right) }{c}\right) }{N\exp \left( -N-N\frac{e^{\left( 1-cj\right)
}}{c}\right) } \\
&=&\frac{1}{N}\exp \left( -\left( 1-N\right) +\left( 1-N\right) cj\right) \,.
\end{eqnarray*}
As a consequence, the entanglement entropy of a one-link state is immediately recovered
\begin{eqnarray*}
S_{EE} &\sim &\lim_{N\rightarrow 1}\frac{1}{1-N}\left( -\left( 1-N\right)
+\left( 1-N\right) cj - \ln N\right)  \\
&=&cj\,,
\end{eqnarray*}
which is a result that implies the area law. 

\section{Entanglement in many-link states} \label{ap.A}
\noindent
For many-link states, we can perform the same analysis deployed in the previous section. Again, we first focus on the normalization of the state:
\begin{eqnarray*}
I &=&\int dh\prod_{l}d_{j_{l}}\int dn_{l}\langle j_{l},j_{l}|n_{l}^{\dag
}e^{-cL_{g_{l}}-\frac{\exp \left( 1-cL_{g_{l}}\right) }{c}
}hn_{l}|j_{l},j_{l}\rangle  \\
&=&\int dh\prod_{l}d_{j_{l}}\int dn_{l}e^{S_{l}} \\
&=&\int dh\prod_{l}d_{j_{l}}\sum_{k_{l}}^{\infty }\frac{\left( -\right)
^{k_{l}}e^{k_{l}}}{k_{l}!c^{k_{l}}}\int dn_{l}e^{S_{k_{l}}^{l}}\,,
\end{eqnarray*}
where
\begin{equation*}
S=\sum_{l}\ln \langle j_{l},j_{l}|n_{l}^{\dag }e^{-cL_{g_{l}}-\frac{\exp
\left( 1-cL_{g_{l}}\right) }{c}}hn_{l}|j_{l},j_{l}\rangle \,.
\end{equation*}
In the semi-classical limit, the saddle point solution which gives the
leading contribution is expressed by 
\begin{equation*}
h=1,\quad n_{l}=g_{l},\quad \sum_{l}j_{l}\vec{n}_{l}=0\,.
\end{equation*}
The leading contribution of the action $S_{l}$ in the saddle point is found to be
\begin{equation*}
S_{l}\sim -cj_{l}-\frac{\exp \left( 1-cj_{l}\right) }{c}\,.
\end{equation*}
The Hessian can also be calculated explicitly. Using the parameterization $h=\exp \left( i\theta _{a}L^{a}\right) $, with $a=1,2,3$, we find three
kinds of components for the Hessian,
\begin{equation*}
-\frac{1}{2}H=
\begin{pmatrix}
A & B \\ 
B^{T} & C
\end{pmatrix}\,,
\end{equation*}
where  
\begin{equation*}
A\equiv -\frac{1}{2}\delta _{n_{l}}^{2}S|\sim 2j_{l}\bar{\eta}_{l}\eta
_{l}\left( -c+\exp \left( 1-cj_{l}\right) \right) \,,
\end{equation*}
\begin{eqnarray*}
&B\equiv \delta _{\theta _{a}}\delta _{n_{l}}S\\
&=\!-i\bar{\eta}_{l}\theta
_{a}e^{c-\exp ^{1-cj}}\!\!j_{l}\langle \downarrow |n_{l}^{\dag
}L^{a}n_{l}|\uparrow \rangle \!-\!i\eta _{l}\theta _{a}j_{l}\langle \uparrow
|n_{l}^{\dag }L^{a}n_{l}|\downarrow \rangle \,,
\end{eqnarray*}
\begin{equation*}
C\equiv \delta _{\theta _{b}}\delta _{\theta _{a}}S=\sum_{l}\theta
_{a}\theta _{b}\frac{j_{l}}{4}\left( \left( \delta _{ba}+i\epsilon
_{bac}n_{l}^{c}\right) -n_{l}^{b}n_{l}^{a}\right) \,.
\end{equation*}
For the third one of these latter relations, when $a=b$ we find 
\begin{equation*}
\delta _{\theta _{a}}^{2}S=\sum_{l}\theta _{a}^{2}\frac{j_{l}}{4}\left(
1-\left( n_{l}^{a}\right) ^{2}\right) \,,
\end{equation*}
while when $a\neq b$ we recover
\begin{eqnarray*}
\delta _{\theta _{b}}\delta _{\theta _{a}}S &=&\sum_{l}\theta _{a}\theta _{b}
\frac{j_{l}}{4}\left( i\epsilon _{bac}n_{l}^{c}-n_{l}^{b}n_{l}^{a}\right)  \\
&\equiv &\sum_{l}\theta _{a}\theta _{b}j_{l}C_{ab}\,.
\end{eqnarray*}
The determinant for the Hessian can be recast as 
\begin{equation*}
\det \left( -\frac{1}{2}H\right) =\det A\det \left( C-B^{T}A^{-1}B\right) \,.
\end{equation*}
It is straightforward to convince ourselves that 
\begin{equation*}
\left( \det A\right) ^{-1/2}\sim \prod_{l}j_{l}\exp \left( 1-cj_{l}\right) \,,
\end{equation*}%
and then that
\begin{equation*}
\left( B^{T}A^{-1}B\right) _{ab}\sim \sum_{l}-j_{l}e^{cj_{l}}\langle
\downarrow |n_{l}^{\dag }L^{a}n_{l}|\uparrow \rangle \langle \uparrow
|n_{l}^{\dag }L^{b}n_{l}|\downarrow \rangle \,.
\end{equation*}
Furthermore, an easy computation shows that
\begin{eqnarray*}
\langle  &\uparrow &|n^{\dag }L^{a}n|\downarrow \rangle \langle \downarrow
|n^{\dag }L^{b}n|\uparrow \rangle  \\
&=&\langle \uparrow |n^{\dag }L^{a}L^{b}n|\uparrow \rangle -\langle \uparrow
|n^{\dag }L^{a}n|\uparrow \rangle \langle \uparrow |n^{\dag }L^{b}n|\uparrow
\rangle  \\
&=&\frac{1}{4}\left( \left( \delta _{ab}+i\epsilon _{abc}n^{c}\right)
-n^{a}n^{b}\right)  \\
&=&%
\Bigg\{
\begin{array}{c}
\frac{1}{4}\left( 1-\left( n^{a}\right) ^{2}\right) ,\qquad {\rm for}\quad a=b\,, \\ 
\frac{1}{4}\left( i\epsilon _{abc}n^{c}-n^{a}n^{b}\right), \qquad {\rm for}\quad a\neq b \,.
\end{array}%
\end{eqnarray*}%
As a consequence, we can write
\begin{equation*}
\left( B^{T}A^{-1}B\right) _{ab}\sim \sum_{l}-j_{l}e^{cj_{l}}C_{ab}^{l}\,,
\end{equation*}%
from which it follows that
\begin{eqnarray*}
&&\det \left( C-B^{T}A^{-1}B\right)  \\
&=&\det \left( \sum_{l}j_{l}\left( 1-e^{cj_{l}}\right) C_{ab}^{l}\right)  \\
&=&\det \left( \sum_{l}j_{l}\left( 1-e^{cj_{l}}\right) C_{ab}^{l}\right)  \\
&\sim &j^{3}e^{3cj}\,,
\end{eqnarray*}
$j$ denoting the average of $j_{l}$. Then the leading term of the
normalization is finally recovered to be
\begin{equation*}
I\sim \frac{1}{j^{3/2}e^{3cj/2}}\prod_{l}\exp \left( -1-\frac{\exp \left(
1-cj_{l}\right) }{c}\right) \,.
\end{equation*}
Using again the replica trick, we can calculate the entropy of the reduced density matrix, the latter being
\begin{eqnarray*}
\rho _{\left\{ j\right\} }&=&\frac{1}{I}\int
dh\prod_{l}\sum_{m_{l}n_{l}}D_{m_{l}n_{l}}^{j_{l}}\left( e^{-cL_{g_{l}}-%
\frac{\exp \left( 1-cL_{g_{l}}\right) }{c}}h\right) \\
&\phantom{a}&\times |j_{l},m_{l}\rangle \langle j_{l},n_{l}| \,.
\end{eqnarray*}
As done in the previous section, the trace of $N$ copies of $\rho _{\left\{
j\right\} }$ is expressed as
\begin{eqnarray*}
tr\rho _{\left\{ j\right\} }^{N} \!&=&\!\frac{1}{I^{N}}\int
\prod_{s}^{N}dh_{s}\prod_{l}\chi ^{j_{l}}\!\!\left( \prod_{s}^{N} \!\! \left(
e^{-cL_{g_{l}}^{s}-\frac{\exp \left( 1-cL_{g_{l}}^{s}\right) }{c}
}h_{s}\right)\!\! \right)  \\
&\equiv &\frac{1}{I^{N}}\prod_{l}d_{j_{l}}^{N}\int \prod_{s}^{N}dh_{s}\int
dn_{l}^{s}e^{S}\,,
\end{eqnarray*}
where
\begin{equation*}
S=\sum_{l}\sum_{s}\ln \langle j_{l},j_{l}|n_{l}^{s\dag }e^{-cL_{g_{l}}-\frac{
\exp \left( 1-cL_{g_{l}}\right) }{c}}h^{s}n_{l}^{s+1}|j_{l},j_{l}\rangle \,.
\end{equation*}
The saddle point solution of the ``action" is read out of the relations 
\begin{equation*}
h_{s}=1,\quad n_{l}^{s}=g_{l},\quad \sum_{l}j_{l}\vec{n}_{l}^{s}=0\,.
\end{equation*}
The leading term of the action is
\begin{equation*}
S=\sum_{l}-Ncj_{l}-N\frac{\exp \left( 1-cj_{l}\right) }{c}\,.
\end{equation*}
The components of Hessian $H$ are
\begin{equation*}
\delta _{n_{l}^{s}}^{2}S=-4j_{l}\eta _{l}^{s}\bar{\eta}_{l}^{s},
\end{equation*}%
\begin{equation*}
\delta _{n_{l}^{s+1}}\delta _{n_{l}^{s}}S\sim 2j_{l}\bar{\eta}_{l}^{s}\eta
_{l}^{s+1}e^{c-\exp \left( 1-cj_{l}\right) },
\end{equation*}%
\begin{equation*}
\delta _{\theta _{a}^{s}}\delta _{n_{l}^{s}}S=i\theta _{a}^{s}\bar{\eta}%
_{l}^{s}e^{c-\exp ^{1-cj_{l}}}2j_{l}\langle \downarrow |n_{l}^{\dag
}L_{s}^{a}n_{l}|\uparrow \rangle ,
\end{equation*}
\begin{equation*}
\delta _{\theta _{a}^{s-1}}\delta _{n_{l}^{s}}S=i\theta _{a}^{s-1}\eta
_{l}^{s}2j_{l}\langle \uparrow |n_{l}^{\dag }L_{s-1}^{a}n_{l}|\downarrow
\rangle ,
\end{equation*}
\begin{equation*}
\delta _{\theta _{a}^{s}}^{2}S=-\sum_{l}\theta _{b}\theta _{a}\frac{j_{l}}{2}
\left( \left( \delta _{ba}+i\epsilon _{bac}n_{l}^{c}\right)
-n_{l}^{b}n_{l}^{a}\right) \,.
\end{equation*}
As we have noticed for the normalization part, the Hessian too has three components, which we denote as $A$, $B$ and $C$. $A$ contains $\delta _{n_{l}^{s}}^{2}S$ and $\delta
_{n_{l}^{s+1}}\delta _{n_{l}^{s}}S$, $B$ contains $\delta _{\theta
_{a}^{s}}\delta _{n_{l}^{s}}S$ and $\delta _{\theta _{a}^{s-1}}\delta
_{n_{l}^{s}}S$, while $\delta _{\theta _{a}^{s}}^{2}S$ enters $C$. The determinant of the Hessian is then cast as
\begin{equation*}
\det H\sim \det A\det \left( C-B^{T}A^{-1}B\right) .
\end{equation*}
The determinant of $A$ is found to be
\begin{equation*}
\det A=\prod_{l}\left( j_{l}\exp \left( 1-cj_{l}\right) \right) ^{2N}.
\end{equation*}%
To compute the determinant of $C-B^{T}A^{-1}B$ is highly non-trivial.
However remind that the entanglement entropy counts the number of
microstates $\Omega \left( j_{l}\right) $ on the boundary, and that the gauge
invariance introduces a reduction on $\Omega \left( j_{l}\right) $ by a
factor $\left( \sum_{l}j_{l}^{2}\right) ^{-3/2}$ \cite{bianchi2011black}. This entails a logarithmic correction to the entanglement entropy. So finally, for a $L$-link state, the entanglement
entropy is recovered to be     
\begin{equation*}
S_{EE}=\sum_{l}^{L}cj_{l} -\frac{3}{2}\ln \left( \sum_{l}j_{l}^{2}\right) \,.
\end{equation*}

\end{document}